\documentclass{aipproc}
\layoutstyle{6x9}
\begin{document}

\title{Strong one-pion decay of ground state charmed baryons}
\author{C. Albertus}{address={Departamento de F\'{\i}sica Moderna, Universidad de
Granada, E-18071 Granada, Spain.}}
\author{ E. Hern\'andez}{address={Grupo de F\'\i sica Nuclear, Departamento de
F\'\i sica Fundamental, 
Facultad de Ciencias, E-37008 Salamanca, Spain.}}  
\author {J. Nieves}{address={Departamento de F\'{\i}sica Moderna, Universidad de
Granada, E-18071 Granada, Spain.}}
\author{ J. M. Verde-Velasco}{address={Grupo de F\'\i sica Nuclear, Departamento de
F\'\i sica Fundamental, 
Facultad de Ciencias, E-37008 Salamanca, Spain.}} 



\begin{abstract}
We evaluate the  widths for the strong decays
$\Sigma_c\to\Lambda_c\,\pi$, $\Sigma_c^*\to\Lambda_c\,\pi$ and 
$\Xi_c^{*}\to\Xi_c\,\pi$. The calculations have been done within a
nonrelativistic constituent quark model with wave functions that take
advantage of the constraints imposed by heavy quark symmetry. Partial
conservation of axial current hypothesis
allows us to determine the strong vertices from an analysis of the
weak axial current matrix elements. Our results are in good agreement
with experimental data.
\end{abstract}

\maketitle


We present results for the strong widths  corresponding to the 
$\Sigma_c\to\Lambda_c\,\pi$, $\Sigma_c^*\to\Lambda_c\,\pi$ and 
$\Xi_c^{*}\to\Xi_c\,\pi$ decays, evaluated in  a
nonrelativistic constituent quark model (NRCQM).
The baryon wave functions used in this calculation 
were obtained in Ref.~\cite{albertus04} solving the three--body problem with a simple variational
ansazt made possible  by Heavy Quark Symmetry (HQS) constraints. 
In order to check the sensitivity of the results to the quark--quark interaction
we use five different  quark-quark potentials taken from
 Refs.~\cite{silvestre96,bhaduri81}. All of them   include a
confining term plus Coulomb and hyperfine terms coming from a one--gluon
exchange potential. We evaluate the pion emission amplitude
 using a one--quark pion emission model in which, through partial conservation of
 axial current (PCAC), we relate that amplitude to  weak current matrix 
 elements, the ones we evaluate.
To the best of our knowledge 
this is the first fully dynamical calculation of these 
observables  done within a nonrelativistic approach.
The work reported in this contribution is now 
published in Ref.~\cite{nuevo} and interested readers can find there full
details of the calculation.  


\begin{table}[h!!!!!!!!]
\begin{tabular}{l| c c c}
\hline
   &
 \hspace{.25cm}$\Gamma(\Sigma_c^{++}\to\Lambda_c^+\pi^+)$ \hspace{.25cm} &
 \hspace{.25cm} $\Gamma(\Sigma_c^{+}\to\Lambda_c^+\pi^0)$ \hspace{.25cm} &
 \hspace{.25cm}$\Gamma(\Sigma_c^{0}\to\Lambda_c^+\pi^-)$ \hspace{.25cm} 
 \\
 &[MeV]&[MeV]&[MeV]\\
\hline

This work 
 &$2.41\pm0.07\pm0.02$ &$2.79\pm0.08\pm0.02$ 
&$2.37\pm0.07\pm0.02$\\
\hline
Experiment   & $2.3\pm0.2\pm 0.3$~\cite{cleo02} & $< 4.6$ (C.L.=90\%)~\cite{cleo01} & 
$2.5\pm0.2\pm0.3$~\cite{cleo02} \\
  &$2.05^{+0.41}_{-0.38}\pm0.38$~\cite{focus02} & &
 $1.55^{+0.41}_{-0.37}\pm0.38$~\cite{focus02}\\
 \hline
Theory  &  & &\\
CQM
&  $1.31\pm 0.04$~\cite{rosner95} &$1.31\pm 0.04$~\cite{rosner95}& 
$1.31\pm 0.04$~\cite{rosner95}\\
 & $2.025^{+1.134}_{-0.987}$~\cite{pirjol97} & &
 $1.939^{+1.114}_{-0.954}$~\cite{pirjol97} \\
HHCPT  & 2.47, 4.38~\cite{yan92}& 2.85,
5.06~\cite{yan92}&2.45, 4.35~\cite{yan92}\\
 &  2.5~\cite{huang95}&  3.2~\cite{huang95}& 2.4~\cite{huang95}\\
 &  &  &$1.94\pm0.57$~\cite{cheng97}\\
LFQM &  1.64 ~\cite{tawfiq98}&1.70 ~\cite{tawfiq98} & 1.57 ~\cite{tawfiq98}\\
RTQM & $2.85\pm0.19$~\cite{ivanov9899} &$3.63\pm0.27$~\cite{ivanov9899}
  & $2.65\pm0.19$~\cite{ivanov9899} \\
\hline
\end{tabular}
\caption{ 
Total widths $\Gamma(\Sigma_c^{++}\to\Lambda_c^+\pi^+)$,
$\Gamma(\Sigma_c^{+}\to\Lambda_c^+\pi^0)$ and
$\Gamma(\Sigma_c^{0}\to\Lambda_c^+\pi^-)$.
}
\label{tab:siglam}
\end{table}

In Table \ref{tab:siglam} we present our results for 
the widths 
$\Gamma(\Sigma_c^{++}\to\Lambda_c^+\pi^+)$,
$\Gamma(\Sigma_c^{+}\to\Lambda_c^+\pi^0)$ and
$\Gamma(\Sigma_c^{0}\to\Lambda_c^+\pi^-)$. 
They show two classes of errors. The
first one  reflects the variation of our results with the 
 quark--quark potential used,
while  the second one is of numerical origin.
Our results are in  very good agreement with the experimental data 
measured by the CLEO 
Collaboration \cite{cleo02,cleo01} and also in  reasonable agreement with 
the data
from the FOCUS Collaboration \cite{focus02}. 
We also show  theoretical results obtained in  different approaches
like  the constituent
quark model (CQM),  heavy hadron chiral perturbation theory
(HHCPT), and in  relativistic quark models like the light-front quark model
(LFQM) and the relativistic three-quark model (RTQM).  


Results for 
the  total widths $\Gamma(\Sigma_c^{*\,++}\to\Lambda_c^+\pi^+)$, 
$\Gamma(\Sigma_c^{*\,+}\to\Lambda_c^+\pi^0)$ and
$\Gamma(\Sigma_c^{*\,0}\to\Lambda_c^+\pi^-)$
appear in Table~\ref{tab:sig*lam}. 
Our central value for $\Gamma(\Sigma_c^{*\,++}\to\Lambda_c^+\pi^+)$ is above 
the central value for the
latest experimental determination by the CLEO Collaboration in 
Ref.~\cite{cleo05}. For some of the potentials that we use, the AP1 and AP2 
potentials of
Ref.~\cite{silvestre96}, our results  
are within experimental errors. The central value for
$\Gamma(\Sigma_c^{*\,+}\to\Lambda_c^+\pi^0)$ is
slightly above the upper experimental bound determined  by the 
CLEO Collaboration in Ref.~\cite{cleo01}, but again, we obtain results 
which are below the experimental bound for the above mentioned potentials.
Finally  for  $\Gamma(\Sigma_c^{*\,0}\to\Lambda_c^+\pi^-)$
we get a nice agreement with experiment.\\

\begin{table}[h!!!]
\begin{tabular}{l| c c c}
\hline
 & \hspace{.25cm}$\Gamma(\Sigma_c^{*\,++}\to\Lambda_c^+\pi^+)$ \hspace{.25cm}
 & \hspace{.25cm}$\Gamma(\Sigma_c^{*\,+}\to\Lambda_c^+\pi^0)$ \hspace{.25cm}
 & \hspace{.25cm}$\Gamma(\Sigma_c^{*\,0}\to\Lambda_c^+\pi^-)$ \hspace{.25cm}\\
 &[MeV]&[MeV]&[MeV]\\
\hline

This work 
&$17.52\pm0.74\pm0.12$&$17.31\pm0.73\pm0.12$ &
$16.90\pm0.71\pm0.12$\\
\hline
Experiment   & $14.1^{+1.6}_{-1.5}\pm 1.4$~\cite{cleo05}& $< 17$ (C.L.=90\%)~\cite{cleo01} & $16.6^{+1.9}_{-1.7}\pm1.4$~\cite{cleo05}\\
\hline
Theory &&&\\
CQM & 20~\cite{rosner95}&20~\cite{rosner95} &20~\cite{rosner95}\\
HHCPT    & 25~\cite{huang95}&25~\cite{huang95} &25~\cite{huang95}\\
LFQM    & 12.84~\cite{tawfiq98} & & 12.40~\cite{tawfiq98}\\
RTQM& $21.99\pm0.87$~\cite{ivanov9899} &
  & $21.21\pm0.81$~\cite{ivanov9899} \\
\hline
\end{tabular}
\caption{ 
Total widths $\Gamma(\Sigma_c^{*\,++}\to\Lambda_c^+\pi^+)$,
$\Gamma(\Sigma_c^{*\,+}\to\Lambda_c^+\pi^0)$ and
$\Gamma(\Sigma_c^{*\,0}\to\Lambda_c^+\pi^-)$. 
}
\label{tab:sig*lam}
\end{table}\vspace{1cm}

Finally in Table \ref{tab:cas*cas} we present our results for
the widths $\Gamma(\Xi_c^{*\,+}\to\Xi_c^0\pi^+)$,
$\Gamma(\Xi_c^{*\,+}\to\Xi_c^+\pi^0)$,
$\Gamma(\Xi_c^{*\,0}\to\Xi_c^+\pi^-)$ and
$\Gamma(\Xi_c^{*\,0}\to\Xi_c^0\pi^0)$,  and the total
widths  
$\Gamma(\Xi_c^{*\,+}\to\Xi_c^0\pi^++\Xi_c^+\pi^0)$  and
$\Gamma(\Xi_c^{*\,0}\to\Xi_c^0\pi^0+\Xi_c^+\pi^-)$. 
Our central value for $\Gamma(\Xi_c^{*\,+}\to\Xi_c^0\pi^++\Xi_c^+\pi^0)$ 
is slightly
above the experimental bound obtained by the CLEO Collaboration \cite{cleo96}, 
being below that bound for the AP1 and AP2 potentials. For  
$\Gamma(\Xi_c^{*\,0}\to\Xi_c^+\pi^-+\Xi_c^0\pi^0)$ our result
is well below the CLEO Collaboration experimental bound in
Ref.\cite{cleo95}. \\
\begin{table}[h!!!!!]
\begin{tabular}{l|c c c c c}
\hline 
 &\hspace{.01cm} $\Gamma(\Xi_c^{*\,+}\to\Xi_c^0\pi^+)$ \hspace{.01cm} 
 &\hspace{.01cm} $\Gamma(\Xi_c^{*\,+}\to\Xi_c^+\pi^0)$ \hspace{.01cm} 
 &\hspace{.01cm} $\Gamma(\Xi_c^{*\,0}\to\Xi_c^+\pi^-)$ \hspace{.01cm}
 &\hspace{.01cm} $\Gamma(\Xi_c^{*\,0}\to\Xi_c^0\pi^0)$ \hspace{.01cm} \\
 &[MeV] & [MeV]&[MeV]&[MeV]\\ \hline

This work  
&$1.84\pm0.06\pm0.01$&
$1.34\pm0.04\pm0.01$&$2.07\pm0.07\pm0.01$&$0.956\pm0.030\pm0.007$\\
\hline
Theory  &&&\\
LFQM   &$1.12$~\cite{tawfiq98} & $0.69$~\cite{tawfiq98} &
  $1.16$~\cite{tawfiq98}& $0.72$~\cite{tawfiq98}\\
RTQM &$1.78\pm0.33$~\cite{ivanov9899} &$1.26\pm0.17$~\cite{ivanov9899}&$2.11\pm0.29$~\cite{ivanov9899}
  & $1.01\pm0.15$~\cite{ivanov9899} \\
\hline
\multicolumn{1}{l|}{}&\multicolumn{2}{c}
{$\Gamma(\Xi_c^{*\,+}\to\Xi_c^0\pi^++\Xi_c^+\pi^0)$}
&\multicolumn{2}{c}{$\Gamma(\Xi_c^{*\,+}\to\Xi_c^0\pi^++\Xi_c^+\pi^0)$}\\
 &\multicolumn{2}{c}{[MeV]}
&\multicolumn{2}{c}{[MeV]}\\
\hline
\multicolumn{1}{l|}{This work}&\multicolumn{2}{c}{$3.18\pm0.10\pm0.01$}&
\multicolumn{2}{c}{$3.03\pm0.10\pm0.01$}\\
\hline
\multicolumn{1}{l|}{Experiment} & \multicolumn{2}{c}
{$<3.1$ (C.L.=90\%)~\cite{cleo96}}&\multicolumn{2}{c}{$<5.5$ 
(C.L.=90\%)~\cite{cleo95}}\\
\hline
Theory  &&&\\
CQM & \multicolumn{2}{c}{$<2.3\pm0.1$~\cite{rosner95}}  &
\multicolumn{2}{c}{$<2.3\pm0.1$~\cite{rosner95}}\\
 &  \multicolumn{2}{c}{1.191\,--\, 3.971~\cite{pirjol97}} & 
 \multicolumn{2}{c}{1.230\, --\, 4.074~\cite{pirjol97}}\\
 HHCPT & \multicolumn{2}{c}{$2.44\pm0.85$~\cite{cheng97}} &
  \multicolumn{2}{c}{$2.51\pm0.88$~\cite{cheng97}}\\
LFQM &  \multicolumn{2}{c}{1.81~\cite{tawfiq98}} &
 \multicolumn{2}{c}{1.88~\cite{tawfiq98}}\\
RTQM & \multicolumn{2}{c}{$3.04\pm0.50$~\cite{ivanov9899}} &
 \multicolumn{2}{c}{$3.12\pm0.33 $~\cite{ivanov9899}}\\
 \hline
\end{tabular}
\caption{ Values for  the 
decay widths $\Gamma(\Xi_c^{*\,+}\to\Xi_c^0\pi^+)$,
$\Gamma(\Xi_c^{*\,+}\to\Xi_c^+\pi^0)$,
$\Gamma(\Xi_c^{*\,0}\to\Xi_c^+\pi^-)$ and
$\Gamma(\Xi_c^{*\,0}\to\Xi_c^0\pi^0)$.
}
\label{tab:cas*cas}
\end{table}

The  results we obtain are stable against the different quark--quark
interactions, with variations in the decay widths at the level of
$6-8\% 
$. They show an overall good agreement with experimental data
for all reactions analyzed. The agreement is, in most cases,  better than the one obtained by
other models.

\section{acknowledgments}
This research was supported by DGI and FEDER funds, under contracts
BFM2002-03218, BFM2003-00856 and  FPA2004-05616,  by the Junta de Andaluc\'\i a and
Junta de Castilla y Le\'on under contracts FQM0225 and
SA104/04, and it is part of the EU
integrated infrastructure initiative
Hadron Physics Project under contract number
RII3-CT-2004-506078. 


\end{document}